\documentclass[a4paper,11pt]{article}

\usepackage[utf8]{inputenc}
\usepackage[english]{babel}
\usepackage{amssymb}
\usepackage{amsmath}
\usepackage{graphicx}
\usepackage{float}
\usepackage{jheppub}
\usepackage[font=footnotesize]{caption}
\usepackage[font=footnotesize]{subcaption}
\usepackage{dsfont} 
\usepackage{appendix}

\makeatletter
\def\@fpheader{\bigskip\relax}
\makeatother

\newcommand{\be}{\begin{equation}}
\newcommand{\ee}{\end{equation}}
\newcommand{\ben}{\begin{equation*}}
\newcommand{\een}{\end{equation*}}
\newcommand{\ba}{\begin{aligned}}
\newcommand{\ea}{\end{aligned}}
\newcommand{\dd}{\,\mathrm{d}}
\newcommand{\lp}{\left(}
\newcommand{\rp}{\right)}

\newcommand{\pd}{\partial}

\newcommand{\bra}[1]{\left<{#1}\right|}
\newcommand{\ket}[1]{\left|{#1}\right>}

\newcommand{\csch}{\,\mathrm{csch}}

\newcommand{\eq}[1]{\begin{align}#1\end{align}}
\newcommand{\omp}{{\nu_+}}
\newcommand{\omm}{{\nu_-}}
\newcommand{\xp}{x_+}
\newcommand{\xm}{x_-}

\title{\boldmath Precursors, Gauge Invariance, and Quantum Error Correction in AdS/CFT}
\author{Ben Freivogel, Robert A. Jefferson, and Laurens Kabir}
\affiliation{ITFA and GRAPPA\\Universiteit van Amsterdam,\\Science Park 904, Amsterdam, the Netherlands}

\emailAdd{benfreivogel@gmail.com}
\emailAdd{rjefferson@uva.nl}
\emailAdd{laurenskabir@gmail.com}

\abstract{A puzzling aspect of the AdS/CFT correspondence is that a single bulk operator can be mapped to multiple different boundary operators, or precursors. By improving upon a recent model of Mintun, Polchinski, and Rosenhaus, we demonstrate explicitly how this ambiguity arises in a simple model of the field theory. In particular, we show how gauge invariance in the boundary theory manifests as a freedom in the smearing function used in the bulk-boundary mapping, and explicitly show how this freedom can be used to localize the precursor in different spatial regions. We also show how the ambiguity can be understood in terms of quantum error correction, by appealing to the entanglement present in the CFT. The concordance of these two approaches suggests that gauge invariance and entanglement in the boundary field theory are intimately connected to the reconstruction of local operators in the dual spacetime.}

\begin{document}
\maketitle 
\flushbottom

\section{Introduction}
In AdS/CFT, much interest has focused on the emergence of the bulk spacetime from boundary CFT data, but a complete understanding of bulk locality remains elusive. The boundary dual of a local bulk field $\Phi$ located a finite distance $z$ away from the boundary has a remarkably simple formula in terms of an integral of the corresponding local CFT operator $\mathcal{O}$ over space and time:
\be
\Phi(t,x,z)=\int\dd x' d t' K\lp t,x,z|x', t'\rp\mathcal{O}\lp x', t'\rp + O(1/{N})\label{eq:smearing}
\ee
where the kernel $K$ is called the \emph{smearing function}. In the cases where $K$ exists and can be computed, its support on the boundary is a measure for what subregion of the boundary stores the information of a given bulk point. The above construction is often referred to as the HKLL construction, following the extensive work by the eponymous authors \cite{HKLL_2005,HKLL_2006,HKLL_2006_2,Kabat_Lifschytz_2013}.

\begin{figure}[h!]
\centering
\includegraphics[width=0.49\textwidth]{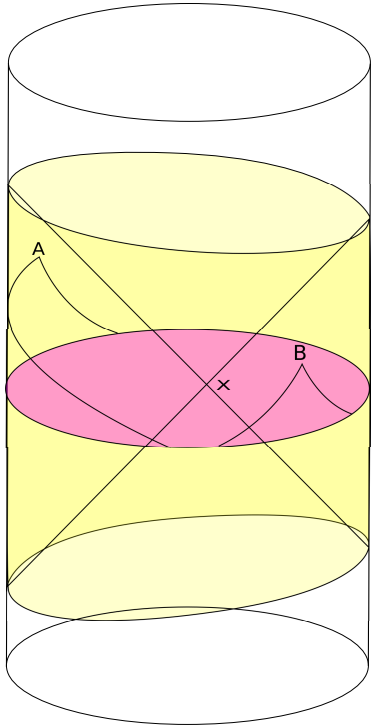}
\caption{Global AdS$_3$, showing the light-cone for a bulk point $x$, which defines a spacelike separated region on the boundary (shaded, yellow online). The corresponding non-local boundary operator is defined \`a la \eqref{eq:smearing} as an integral over this region. The local CFT operators can be time-evolved to a single Cauchy slice (shaded, pink online). This is illustrated schematically for points $A$ and $B$, where we've indicated the null lines on the boundary. In our model, the boundary operators factorize along the light-cone directions, and are trivially evolved to bilocals at the $t=0$ Cauchy slice.\label{fig:cylinder}}
\end{figure}

A perplexing feature of this procedure is that there is a freedom in choosing the smearing function $K$, allowing for a family of different CFT operators corresponding to a given bulk operator. These different CFT operators, when evolved back to one time, can even have support in different spatial regions of the CFT (see figure \ref{fig:cylinder}). We refer to these CFT operators as ``precursors'', because in general they contain information about bulk events before signals from these events have had time to reach the boundary \cite{Polchinski_Susskind_Toumbas_1999,Giddings_Lippert_2001,Freivogel_Giddings_Lippert_2002}. 

 Almheiri, Dong, and Harlow \cite{ADH} pointed out that these different CFT operators cannot be really equal as operators unless the field theory violates the time slice axiom, which is believed to be a fundamental property of physically relevant quantum field theories \cite{Chilian_Fredenhagen_2008}. These authors proposed that the different CFT operators are only equivalent when acting on a certain subclass of states (the ``code subspace''), casting the bulk reconstruction problem in the language of quantum error correction (QEC) and quantum secret sharing \cite{QSS}, in which increasing radial depth into the bulk is interpreted as improved resilience of the boundary theory against local quantum erasures. This idea has been beautifully implemented in several tensor network models \cite{Pastawski_etal_2015, Yang_Hayden_Qi_2015, Hayden_etal_2016}.

Subsequently, Mintun, Polchinski, and Rosenhaus (``MPR'') \cite{MPR} argued that the structure of QEC emerges naturally when one considers the gauge invariance of the boundary field theory. MPR reconcile the representation of a local bulk operator by a number of different CFT operators by pointing out that an operator can be modified by a ``pure-gauge'' contribution that changes its support on the boundary without changing its action on physical states. This suggests that the emergence of local operators in the dual spacetime may be deeply connected with gauge symmetries in the CFT. 

In this work, we clarify the relationship between quantum error correction, gauge freedom, and the localization of precursors, in the context of an explicit bulk reconstruction scheme in AdS$_3$/CFT$_2$. We will first point out a shortcoming of the MPR model: the particular boundary conditions specified by MPR lead to a theory with no bulk dynamics\footnote{We thank Ian Morrison for discussions on this point.}. This difficulty is easily fixed by choosing different boundary conditions, and we revise their model of the CFT accordingly in section \ref{sec:improvement}. We show that with these revisions, the MPR model works as advertised, and provides a nice, tractable model for understanding the CFT encoding of bulk information, including such issues as the role of quantum error correction. 

We show in section \ref{sec:dK} that the known ambiguity in the choice of smearing function arises from the gauge freedom in the $N \to \infty$ limit, and explicitly show how to use this freedom to localize the precursor in different spatial regions. We begin with a standard representation of a local bulk operator spread over the entire CFT, as illustrated in figure \ref{fig:cylinder}, and show that the gauge freedom allows us to localize the precursor within a single boundary Rindler wedge. This result agrees with the claims of MPR, but now in a model with genuine bulk dynamics. We find that this localization procedure works when the bulk field is located inside the corresponding entanglement wedge\footnote{For most of this work, the entanglement and causal wedges agree, and we use the terms interchangeably. We will address the crucial difference between them in section \ref{sec:discussion}.}, consistent with general expectations for bulk reconstruction. This result is independent of the weakly coupled CFT model, and relies only on the freedom in the choice of smearing function.

In section \ref{sec:entanglement}, we instead take a quantum error correction approach to localizing the precursor in a boundary region $A$: we use the entanglement of the ground state to map operators acting on the complement $\bar A$ into operators acting on $A$. We point out how this procedure can fail, and show that it is successful when the above condition is satisfied: the bulk point must lie in the entanglement wedge corresponding to the boundary region under consideration.

Finally, we conclude in section \ref{sec:discussion} with some discussion on the relationship between quantum error correction and gauge freedom in light of our results, and speculate on how our model may be generalized to disconnected boundary regions, where the distinction between the causal and entanglement wedges is significant.

\section{Improved toy model of the bulk-boundary correspondence}\label{sec:improvement}
In this section, we describe the MPR model \cite{MPR} along with our improvements. In the former, the CFT consists of free massless scalars $\phi^i$ in two dimensions, where $i$ is a global $O(N)$ index. The global $O(N)$ symmetry is a simple model for the gauge invariance of the full theory, so ``gauge-invariant'' operators are defined to be operators that are invariant under global $O(N)$ transformations. MPR consider a massless bulk field $\Phi$ in AdS$_3$, and from the two possible consistent quantization schemes \cite{Kleb_Witten_1999},
\be
\Delta_\pm=\frac{d}{2}\pm\sqrt{\frac{d}{2}+m^2}
\ee
they choose boundary conditions such that the bulk field is dual to a $\Delta_-=0$ operator, which they take to be $\phi_i \phi^i$. 

The choice $\Delta=0$ is unfortunate for a number of related reasons. From the CFT point of view, $\Delta = 0$ saturates the unitarity bound. In any dimension, an operator $\mathcal{O}$ saturating the unitarity bound must obey the {\it boundary} equation of motion $\Box \mathcal{O}=0$, meaning that it acts like a free field on the boundary\footnote{We thank Ian Morrison for pointing this out.}. From the bulk point of view, when we impose the boundary condition $\Phi\propto z^{\Delta_-}$ as $z\rightarrow 0$, with $\Delta_- = 0$, there are no solutions to the bulk equation of motion except for the special modes satisfying the boundary wave equation. Therefore, this field does not have true bulk dynamics and is not a good setting to discuss bulk reconstruction; see \cite{Andrade_Marolf_2011} for a more detailed discussion of the $\Delta=0$ limit.

This problem is easily fixed: we simply choose the other boundary condition $\Phi \to z^2$ ($\Delta_+=2$), and take the boundary operator dual to the bulk field to be
\be
\mathcal{O} = \partial_\mu \phi^i \partial^\mu \phi_i ~,
\ee
where the $\phi^i$ are free massless scalar fields as in the MPR model.
Strictly speaking, this is also a poor model for perturbative bulk physics, since the CFT is weakly coupled. However, at the level of two-point functions it suffices to capture the salient features. This improved model is almost identical to \cite{Freivogel_Giddings_Lippert_2002}, which in turn was closely related to \cite{Polchinski_Susskind_Toumbas_1999}.

In the following we suppress the $O(N)$ index $i$ and use light-cone coordinates $x_\pm=t\pm x$ in the boundary, so we can write simply 
\be
\mathcal{O} = \partial_+ \phi\partial_- \phi ~.
\ee
We expand the CFT field $\phi$ in terms of creation and annihilation operators as
\be
{\mathcal \phi}(x_+, x_-) = \int {\dd \nu_+ \over \nu_+} \alpha_{\nu_+} e^{-i \nu_+ x_+} + \int {\dd \nu_- \over \nu_-} \tilde \alpha_{\nu_-} e^{-i \nu_- x_-}~.\label{eq:Pmodes}
\ee
where $\alpha$ and $\tilde \alpha$ correspond to the right and left movers, respectively.
This then yields a simple formula for the ``primary'' operator $\mathcal{O}$,
\be
\mathcal{O}(\xp, \xm)= -\int\dd \omp\dd\omm e^{-i (\nu_+ x_+ + \nu_- \xm)} \alpha_\omp \tilde \alpha_\omm~.
\label{eq:fieldexp}
\ee
In the large-$N$ limit, MPR pointed out that the global $O(N)$ gauge invariance includes the freedom to add to any operator a linear combination of operators of the form $\alpha_{\nu_+} \tilde \alpha_{\nu_-}$ as long as $\nu_+ \nu_- < 0$. 

\section{Localizing the precursor via gauge freedom}\label{sec:dK}
The freedom identified by MPR at first appears distinct from the freedom in the choice of smearing function, but we will show that they are in fact identical. We will then show explicitly how this freedom can be used to localize the precursor within a given boundary region, in an effort to make more precise the role that gauge invariance plays in the localization and non-uniqueness of boundary data.

The precursor for a local bulk field $\Phi$ is defined with support on the entire boundary by eqn. \eqref{eq:smearing}, in the $N\rightarrow\infty$ limit,
\be
\Phi(t,x,z)=\int\dd x'\dd t' K\lp t,x,z|x', t'\rp\mathcal{O}\lp x', t'\rp~.\label{eq:pre}
\ee
The smearing function $K$ in Poincar\'e-AdS$_3$ for a field with conformal dimension $\Delta=2$ is given by \cite{HKLL_2005}
\be
K(t,x,z|t',x')=\log\left(\frac{|z^2+(x-x')^2-(t-t')^2|}{2z} \right)\equiv K. \label{eq:poincareK}
\ee
The ambiguity in the smearing function consists of the freedom to add a function $\delta K$ which in Fourier space satisfies $\nu_+ \nu_- < 0$. This can be understood from the fact that satsifying the bulk wave equation, $\Box\Phi=0$, in global AdS implies that modes with frequency $\omega$ and boundary momentum $\kappa$ satisfying $\omega^2<\kappa^2$ (or in light-cone coordinates, $\nu_+\nu_-<0$) are disallowed. Hence the dual operator has no support on the space of these modes, $\int\dd^2 x ~ \mathcal{O}~ \delta K =0$.

Focusing on a particular Fourier mode, the change in the smearing function is
\be
\delta K(\xp, \xm) = e^{i ( \nu_+ x_+ + \nu_- \xm)}~.
\ee
The corresponding change in the precursor is therefore
\be
\delta \Phi = \int\dd x_+\dd x_- e^{i ( \nu_+ x_+ + \nu_- \xm)} \mathcal{O}~.
\ee
Plugging in the expansion for the field in terms of creation and annihilation operators (\ref{eq:fieldexp}) then gives
\be
\delta \Phi = -\int\dd x_+\dd x_- e^{i ( \nu_+ x_+ + \nu_- \xm)} \int\dd \omp'\dd \omm' e^{-i(\omp' \xp + \omm' \xm)} \alpha_{\omp'} \tilde \alpha_{\omm'}~.
\ee
The spatial integrals can be performed, yielding
\be
\delta \Phi = -\alpha_\omp \tilde \alpha_\omm~.
\ee
This demonstrates that the freedom identified in MPR corresponds precisely to the freedom in the choice of smearing function. In this sense, we will refer to the function $\delta K$ satisfying $\nu_+ \nu_- < 0$ as ``pure gauge'' henceforth.

We are now prepared to investigate the idea of MPR in the context of an explicit HKLL construction \cite{HKLL_2005}, by demonstrating that the gauge freedom can be used to localize the precursor to within a single boundary Rindler wedge. In Poincar\'e light-cone coordinates, the metric for Rindler-AdS$_3$ is
\be
\dd s^2 = {-\dd x_+\dd x_- +\dd z^2 \over z^2}~,
\ee
which naturally leads to a bulk Rindler horizon at $x_+ = x_- = 0$. This horizon defines the bulk Rindler or causal wedge, and our aim is to localize the precursor for a given field in this wedge within the corresponding boundary region. This requires finding the most general pure-gauge function $\delta K$ that we can add such that the new smearing function, $\hat K=K+\delta K$, only has support within that region.

To proceed, we need to know how the pure-gauge mode functions (that is, the Poincar\'e modes with $\nu_+ \nu_- <0$) look in the various Rindler wedges. This can be done by studying the analyticity of the mode functions in the complex plane.

\begin{figure}[h!]
\centering
\includegraphics[width=0.6\textwidth]{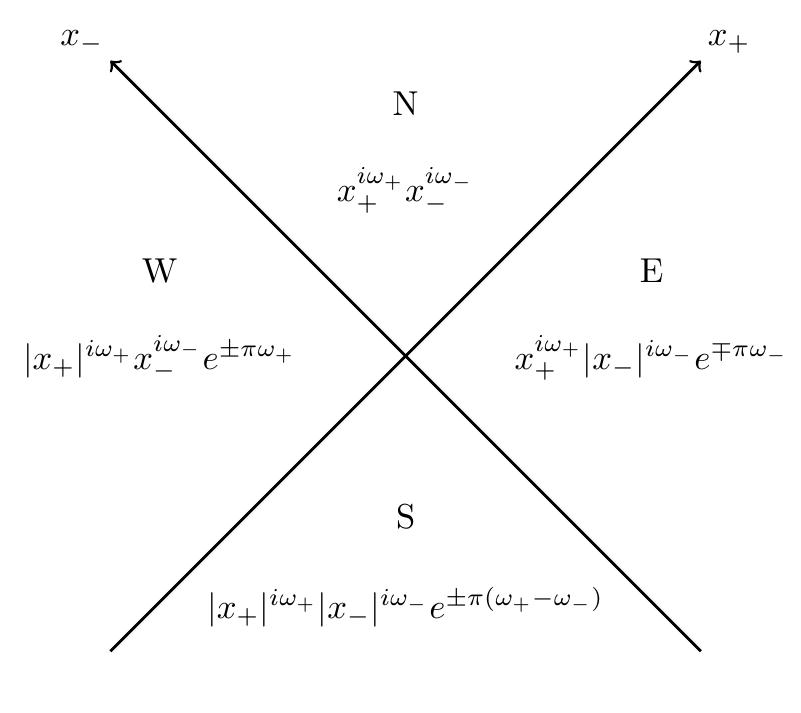}
\caption{Rindler plane in light-cone coordinates, indicating the phase changes in the mode functions \eqref{eq:boundarylimit} when crossing the branch cuts at $x_\pm=0$. The sign choice is arbitrary, but must be consistent across all four quadrants in order to obtain a pure-gauge Poincar\'e mode. We refer to these quadrants throughout as the northern (N), southern (S), eastern (E), and western (W) wedges, labelled in the obvious manner.\label{fig:Axes_sketch}}
\end{figure}

It is convenient to work in terms of the Rindler modes\footnote{In fact, working in terms of the Rindler modes is more than a convenience, it is a necessity, because the resulting smearing function can only be written in Fourier space; it cannot be transformed to position space.}
\be
x_+^{i\omega_+}x_-^{i\omega_-}~.\label{eq:boundarylimit}
\ee
where $x_\pm$ are the Poincar\'e light-cone coordinates, as above. The Rindler plane is sketched in fig. \ref{fig:Axes_sketch}. The above Rindler modes \eqref{eq:boundarylimit} are then defined as-is in the northern quadrant, where $x_+>0$, $x_->0$. We would then like to know what this looks like in the remaining three quadrants. However, getting there requires navigating the branch cuts at $x_+=0$ and/or $x_-=0$. 

Consider moving into the western wedge. We have a choice of contour upon crossing the branch cut at $x_+=0$. Suppose we take the function to be analytic in the lower-half complex $x_+$ plane. Then the transformation from the northern wedge ($x_+>0$) across $x_+=0$ into the western wedge ($x_+<0$), is $x_+\rightarrow|x_+|e^{-i\pi}$, where the minus sign in the exponential corresponds to our choice of contour. The Rindler mode changes as
\be
x_+^{i\omega_+}x_-^{i\omega_-}\rightarrow|x_+|^{i\omega_+}x_-^{i\omega_-}e^{\pi\omega_+}~.
\ee
 Since we chose $x_+$ to be analytic in the lower half-plane, our mode is a superposition of positive frequency Poincar\'e modes $\nu_+>0$.\footnote{Any function $f(x)$ built out of positive frequency Fourier modes (that is $e^{-i \nu x}$ with $\nu>0$) must be analytic in the lower half of the complex $x$-plane, and vice versa.} Had we made the opposite choice for the analyticity of the function, we would take the opposite sign of $\nu_+$. Hence the general transformation across the $x_+=0$ branch cut into the western wedge is
\be
x_+^{i\omega_+}x_-^{i\omega_-}\rightarrow|x_+|^{i\omega_+}x_-^{i\omega_-}e^{\pm\pi\omega_+}\hspace{1cm}\mathrm{(N }\rightarrow\mathrm{ W)}\label{eq:transW}
\ee
where the upper sign is for $\nu_+>0$, lower for $\nu_+<0$. From this relation one immediately writes down the transformation from the northern quadrant across $x_-=0$ into the east ($x_-<0$):
\be
x_+^{i\omega_+}x_-^{i\omega_-}\rightarrow x_+^{i\omega_+}|x_-|^{i\omega_-}e^{\mp\pi\omega_-}\hspace{1cm}\mathrm{(N }\rightarrow\mathrm{ E)}\label{eq:transE}
\ee
where the upper sign is for $\nu_-<0$, lower for $\nu_->0$. Similarly, the transformation of the precursor into the southern quadrant, with two branch crossings, is
\be
x_+^{i\omega_+}x_-^{i\omega_-}\rightarrow 
|x_+|^{i\omega_+}|x_-|^{i\omega_-}e^{\pm\pi\omega_+\mp\pi\omega_-}\hspace{1cm}\mathrm{(N }\rightarrow\mathrm{ S)}~.\label{eq:transS}
\ee
The crucial fact is that the above, with a consistent sign choice (upper or lower), corresponds to a pure-gauge function in Poincar\'e, since we have $\nu_+\nu_-<0$ by construction. The nice feature of this method is that we're guaranteed this without having to explicitly work with Poincar\'e modes, where the meaning of $\nu_+\nu_-<0$ in the various quadrants is not readily visualized. 

From this analyticity analysis, we can immediately write down the general form of the pure-gauge function $\delta K$:
\be
\delta K=\int\dd\omega_+\dd\omega_-
\lp c_{\omega_+\omega_-}f_{\omega_+\omega_-}^{\mathrm{upper}}
+d_{\omega_+\omega_-}f_{\omega_+\omega_-}^{\mathrm{lower}}\rp\label{eq:dK}
\ee
with, from \eqref{eq:transW}, \eqref{eq:transE}, and \eqref{eq:transS},
\be
\ba
f_{\omega_+\omega_-}&=
x_+^{i\omega_+}x_-^{i\omega_-}\Theta(x_+)\Theta(x_-)
+|x_+|^{i\omega_+}x_-^{i\omega_-}e^{\pm\pi\omega_+}\Theta(-x_+)\Theta(x_-)\\
&+x_+^{i\omega_+}|x_-|^{i\omega_-}e^{\mp\pi\omega_-}\Theta(x_+)\Theta(-x_-)
+|x_+|^{i\omega_+}|x_-|^{i\omega_-}e^{\pm\pi\omega_+\mp\pi\omega_-}\Theta(-x_+)\Theta(-x_-)~.\label{eq:f}
\ea
\ee
The labels ``upper" and ``lower" on the functions $f$ in (\ref{eq:dK}) indicate choosing the upper or lower signs in the exponentials in \eqref{eq:f}, and the coefficients $c$ and $d$ are undetermined functions of the momenta. 

We may extract from this general expression the pure-gauge function in momentum space, $\delta\tilde K$, in each of the four quadrants:
\be
\ba
\delta \tilde K_N&=c + d \\
\delta \tilde K_W&= e^{\pi \omega_+} c + e^{-\pi \omega_+ }d \\
\delta \tilde K_S&=e^{\pi( \omega_+ - \omega_-)} c + e^{-\pi( \omega_+ - \omega_-) }d \\
\delta \tilde K_E&=e^{-\pi \omega_-} c + e^{\pi \omega_- }d \label{eq:constraints1} \\
\ea
\ee
where we have suppressed the $\omega_\pm$ subscripts on $c$ and $d$ to minimize clutter.

In order to localize support for the precursor within a single Rindler wedge, we must choose the coefficients $c$ and $d$ such that $\hat{\tilde{ K}}=\tilde K+\delta \tilde K$ is zero in the other three regions. Let us attempt to localize the precursor in the east, so that only $\hat{\tilde K}_E\neq0$. Then the coefficients must be chosen such that
 \be
\delta \tilde K_N =-\tilde K_N~,\;\;\;
\delta \tilde K_W =-\tilde K_W~,\;\;\;
\mathrm{and}\;\;\;\;
\delta \tilde K_S =-\tilde K_S
\label{eq:constraints2}
\ee
where $\tilde K_X$ with $X\in\{E,N,W,S\}$ is the Fourier transform of the smearing function \eqref{eq:poincareK} in the specified wedge,
\be
\tilde K_X\lp\omega_+,\omega_-\rp \equiv \iint_X\dd x_+\dd x_- K(0,a,z|x_+,x_-) |x_+|^{-i \omega_+-1}|x_-|^{-i \omega_- -1}~.\label{eq:FourierK}
\ee
We have  chosen the bulk field to be located at time $t=0$, radial coordinate $z$, and a distance $a$ into the eastern wedge of the bulk.

At a glance, the system \eqref{eq:constraints2} appears overdetermined, as we have three equations and only two unknowns, $c$ and $d$. However, we shall find that the system does indeed have a consistent solution, provided that the bulk point lies within the bulk extension (the causal or entanglement wedge) of the boundary Rindler wedge in which we attempt to localize the smearing function, in this case the east. We shall return to this requirement below.

In the course of solving this system, we rely on the following relations between the Fourier transforms of the smearing function, which we prove in appendix \ref{app:fourier}:
\be
\ba
\tilde K_N&=\cosh(\pi\omega_+)\tilde K_W\\
\tilde K_S&=\cosh(\pi\omega_-)\tilde K_W \label{eq:Ben}\\
\tilde K_E&=\cosh\lp\pi(\omega_+-\omega_-)\rp \tilde K_W
\ea
\ee
Note that the singularities in the smearing function (\ref{eq:poincareK}), which occur when the argument of the logarithm is zero, do not extend into the western quadrant. This is a consequence of the fact that we chose the bulk point to be in the eastern Rindler wedge. The benefit of these relations is that they allow us to rewrite everything in terms of the Fourier transform $\tilde K_W$, which is well-defined.

With the relations \eqref{eq:Ben} in hand, one can show that the system \eqref{eq:constraints2} is solved by
\eq{
c=-\frac{1}{2} e^{-\pi \omega_+}\tilde K_W \qquad d=-\frac{1}{2}e^{\pi \omega_+}\tilde K_W~. \label{eq:cdsolutions}
}
and therefore that the only non-zero portion of the momentum space smearing function, $\hat{\tilde K}_E$, is
\be
\ba
\hat{ \tilde{K}}_E &\equiv \tilde K_E+\delta \tilde K_E = \cosh\lp\pi(\omega_+-\omega_-)\rp \tilde K_W + \left(e^{-\pi \omega_-} c + e^{\pi \omega_- }d\right)\\
&=-2 \sinh(\pi \omega_+) \sinh(\pi \omega_-)\tilde K_W~.\label{eq:KEpre}
\ea
\ee
It then remains to obtain an explicit expression for $\tilde K_W$, which we can do by computing the Fourier transform of \eqref{eq:poincareK} in the western Rindler wedge. The integration is performed in appendix \ref{app:integral}. Substituting the result into \eqref{eq:KEpre}, we have
\be
\hat{\tilde K}_E=-2\pi^2\lp\frac{z}{a}\rp^2a^{-i(\omega_++\omega_-)}{}_2F_1\lp1+i\omega_+,1+i\omega_-,2,\frac{-z^2}{a^2}\rp~,\label{eq:KE}
\ee
which is consistent with results found in the literature \cite{HKLL_2006}. We therefore find that the smeared bulk operator \eqref{eq:pre} at $t=0$, $x=a>0$, and radial distance $z$, with support localized entirely within the eastern Rindler wedge, is given by
\be
\Phi(0,a,z)=-2\pi^2\lp\frac{z}{a}\rp^2\int\dd\omega_+\dd\omega_-a^{-i(\omega_++\omega_-)}{}_2F_1\lp1+i\omega_+,1+i\omega_-,2,\frac{-z^2}{a^2}\rp \tilde{\mathcal{O}}^E_{\omega_+,\omega_-}~,\label{eq:PhiK}
\ee
where $\tilde{\mathcal{O}}^E_{\omega_+,\omega_-}$ is the momentum-space boundary operator, with support in the eastern wedge. We will write this explicitly in Rindler modes (cf. the Poincar\'e expression \eqref{eq:fieldexp}) in the next section, but forgo unnecessary details here. 

The action of the precursor \eqref{eq:PhiK} is UV-sensitive, and only well-defined when acting on an appropriate subclass of states. As we show explicitly in appendix \ref{app:twopoint}, its vacuum two-point function reproduces the correct bulk correlator in the near-horizon limit.

As mentioned previously, a condition on the success of our procedure is that the bulk point be located in the entanglement wedge of the boundary region in which we attempt to localize the precursor. A natural question to ask is whether the gauge freedom in the smearing function can still be used to reconstruct a bulk point located outside the entanglement wedge. As we placed our bulk point in the eastern wedge, this would amount to trying to set the smearing function to zero in the eastern quadrant instead of the western quadrant as we did above. The set of conditions on the pure-gauge function $\delta\tilde K$ is then
 \be
\delta \tilde K_N =-\tilde K_N~,\;\;\;
\delta \tilde K_E =-\tilde K_E~,\;\;\;
\mathrm{and}\;\;\;
\delta \tilde K_S =-\tilde K_S~.
\label{eq:wrongconstraints}
\ee
One can simply check that the overdetermined system of equations \eqref{eq:constraints1}, \eqref{eq:Ben} and \eqref{eq:wrongconstraints} is now inconsistent: there no longer exists a solution for $c$ and $d$. Hence we conclude that our model is consistent with the current understanding of bulk reconstruction, namely that it succeeds when the bulk point is inside -- and fails when the point is outside -- the causal/entanglement wedge. We shall comment more on this in section \ref{sec:discussion}, and elaborate on the distinction between the two types of bulk wedges, but first we turn to an alternative approach of localizing the bulk field, appealing instead to the entanglement structure in the CFT.

\section{Localizing the precursor via entanglement mapping}\label{sec:entanglement}
In this section we will present an alternative method for localizing the precursor. As before, our starting point is the smeared operator in Poincar\'e coordinates \eqref{eq:pre}, which has non-zero support on the entire boundary and can be time-evolved to bilocals at $t=0$. Instead of using the gauge freedom to manipulate the support of the smearing function $K$ however, we will now use entanglement in the field theory to map all bilocal operators into the eastern Rindler wedge. We will explicitly show that this gives the same result as that obtained in the previous section, thereby establishing that the freedom in the smearing function from gauge invariance can be equivalently understood from an entanglement perspective.

In eqn. \eqref{eq:Pmodes}, we expanded the CFT field $\phi$ in terms of Poincar\'e modes. We may equivalently write the mode expansion in terms of Rindler creation ($\omega<0$) and annihilation ($\omega>0$) operators $\beta_{\omega_\pm}$ with left- and right-moving\footnote{In this section, to avoid clutter, we denote right movers by $\beta_{\omega_+}$ and left movers by $\beta_{\omega_-}$, with no tilde on the left movers. Left and right movers commute.} Rindler momenta $\omega_\pm$. These satisfy
\eq{
[\beta_{\omega_\pm},\beta_{\omega'_\pm}]=\omega_{\pm}\delta(\omega_\pm + \omega'_\pm) \qquad \text{and} \qquad \beta_{\omega_\pm}^\dagger=\beta_{-\omega_{\pm}}~.
}
In light-cone coordinates $x_\pm \equiv t\pm x$, the Rindler expansion of the field in the eastern and western wedges (cf. fig. \ref{fig:Axes_sketch}) is, respectively,
\eq{
\phi^E(t,x)&= \int_{-\infty}^{+\infty} \frac{\dd\omega_+}{\omega_+}\; \beta^E_{\omega_+} |x_+|^{-i\omega_+} + \int_{-\infty}^{+\infty} \frac{\dd\omega_-}{\omega_-}\; \beta^E_{\omega_-} |x_-|^{i\omega_-} 
}
\eq{
\phi^W(t,x)&= \int_{-\infty}^{+\infty} \frac{\dd\omega_+}{\omega_+}\; \beta^W_{\omega_+} |x_+|^{i\omega_+} + \int_{-\infty}^{+\infty} \frac{\dd\omega_-}{\omega_-}\; \beta^W_{\omega_-} |x_-|^{-i\omega_-}~,
}
where the Rindler mode functions are chosen such that they are positive frequency with respect to Rindler time, which we take to run upwards in both the eastern and western wedge. The light-cone derivatives are 
\eq{
\partial_+ \phi^E =-i \int_{-\infty}^{+\infty} {\dd\omega_+}\beta^E_{\omega_+} |x_+|^{{-i\omega_+}-1} \qquad \partial_- \phi^E =i \int_{-\infty}^{+\infty} {\dd\omega_-}\beta^E_{\omega_-} |x_-|^{{i\omega_-}-1}\label{eq:derivE}
}
\eq{
\partial_+ \phi^W =i \int_{-\infty}^{+\infty} {\dd\omega_+}\beta^W_{\omega_+} |x_+|^{{i\omega_+}-1} \qquad \partial_- \phi^W =-i \int_{-\infty}^{+\infty} {\dd\omega_-}\beta^W_{\omega_-} |x_-|^{{-i\omega_-}-1}\label{eq:derivW}
}
and are manifestly purely left/right-moving. As a consequence, their time evolution becomes trivial:
\eq{
\partial_+ \phi(t,x)=\partial_+\phi(0,x+t) \qquad \partial_- \phi(t,x)=\partial_-\phi(0,x-t)~.
}
This was to be expected, since $\phi$ satisfies the $1+1$-dimensional wave equation $\Box\phi=0$. This factorization along the null directions allows us to write the precursor, for a bulk operator shifted a distance $a$ into the east, as a bilocal at $t=0$:
\eq{
\Phi(t=0,x=a>0,z)= \int\dd x_+\dd x_- K(0,a,z|x_+,x_-) \partial_+ \phi(0,x_+) \partial_- \phi(0,-x_-)~,\label{eq:PLC}
}
where the smearing function \eqref{eq:poincareK}, in light-cone coordinates, is
\eq{
K(0,a,z|x_+,x_-)=\log\left(\frac{|z^2-(x_+-a)(x_- + a)|}{2z} \right)\equiv K~.\label{eq:poincareKLC}
}
Using \eqref{eq:derivE} and \eqref{eq:derivW}, we can explicitly decompose the integral \eqref{eq:PLC} over all four wedges:
\be
\ba
\Phi(0,a,z)=&-\iint_N\dd x_+\dd x_- \iint \dd\omega_+\dd\omega_- K |x_+|^{{-i\omega_+}-1} |x_-|^{{-i\omega_-}-1} \beta^E_{\omega_+}\beta^W_{\omega_-}\\
&-\iint_S\dd x_+\dd x_-\iint\dd\omega_+\dd\omega_- K |x_+|^{{i\omega_+}-1} |x_-|^{{i\omega_-}-1} \beta^E_{\omega_-}\beta^W_{\omega_+}\\
&+\iint_E\dd x_+\dd x_- \iint\dd\omega_+\dd\omega_- K |x_+|^{{-i\omega_+}-1} |x_-|^{{i\omega_-}-1} \beta^E_{\omega_-}\beta^E_{\omega_+}\\
&+\iint_W\dd x_+\dd x_- \iint\dd\omega_+\dd \omega_- K |x_+|^{{i\omega_+}-1} |x_-|^{{-i\omega_-}-1} \beta^W_{\omega_-}\beta^W_{\omega_+}~.
\ea
\ee
We may write this more succinctly in terms of the Fourier transform \eqref{eq:FourierK}, paying careful attention to the signs of the momenta:
\be
\ba
\Phi(0,a,z)=\iint\dd\omega_+\dd\omega_-&\left(-\tilde K_N\lp\omega_+,\omega_-\rp \beta^E_{\omega_+}\beta^W_{\omega_-} - \tilde K_S\lp-\omega_+,-\omega_-\rp\beta^W_{\omega_+}\beta^E_{\omega_-}\right.\\
&\left.+\tilde K_E\lp\omega_+,-\omega_-\rp\beta^E_{\omega_+}\beta^E_{\omega_-}+ \tilde K_W\lp-\omega_+,\omega_-\rp\beta^W_{\omega_+}\beta^W_{\omega_-}\right)~.
\label{eq:Laurensbilocal}
\ea
\ee

\subsection{Mapping the precursor into the eastern Rindler wedge}
\begin{figure}[h!]
\centering
\includegraphics[width=0.35\textwidth]{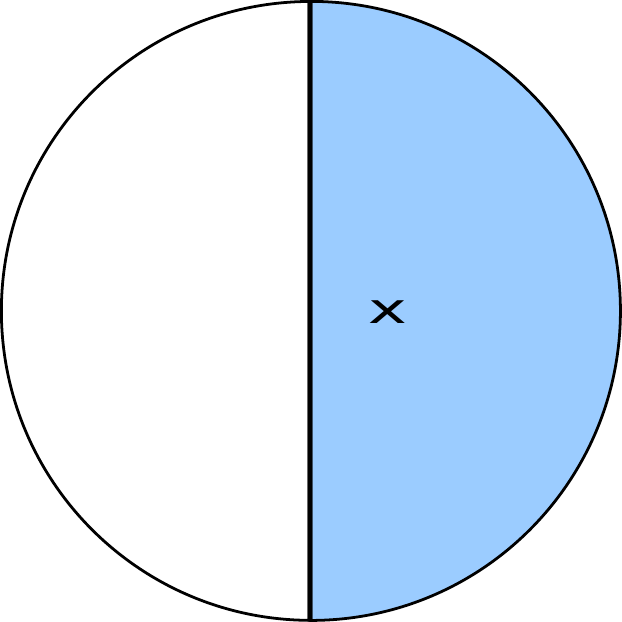}
\hspace{1 cm}
\includegraphics[width=0.4\textwidth]{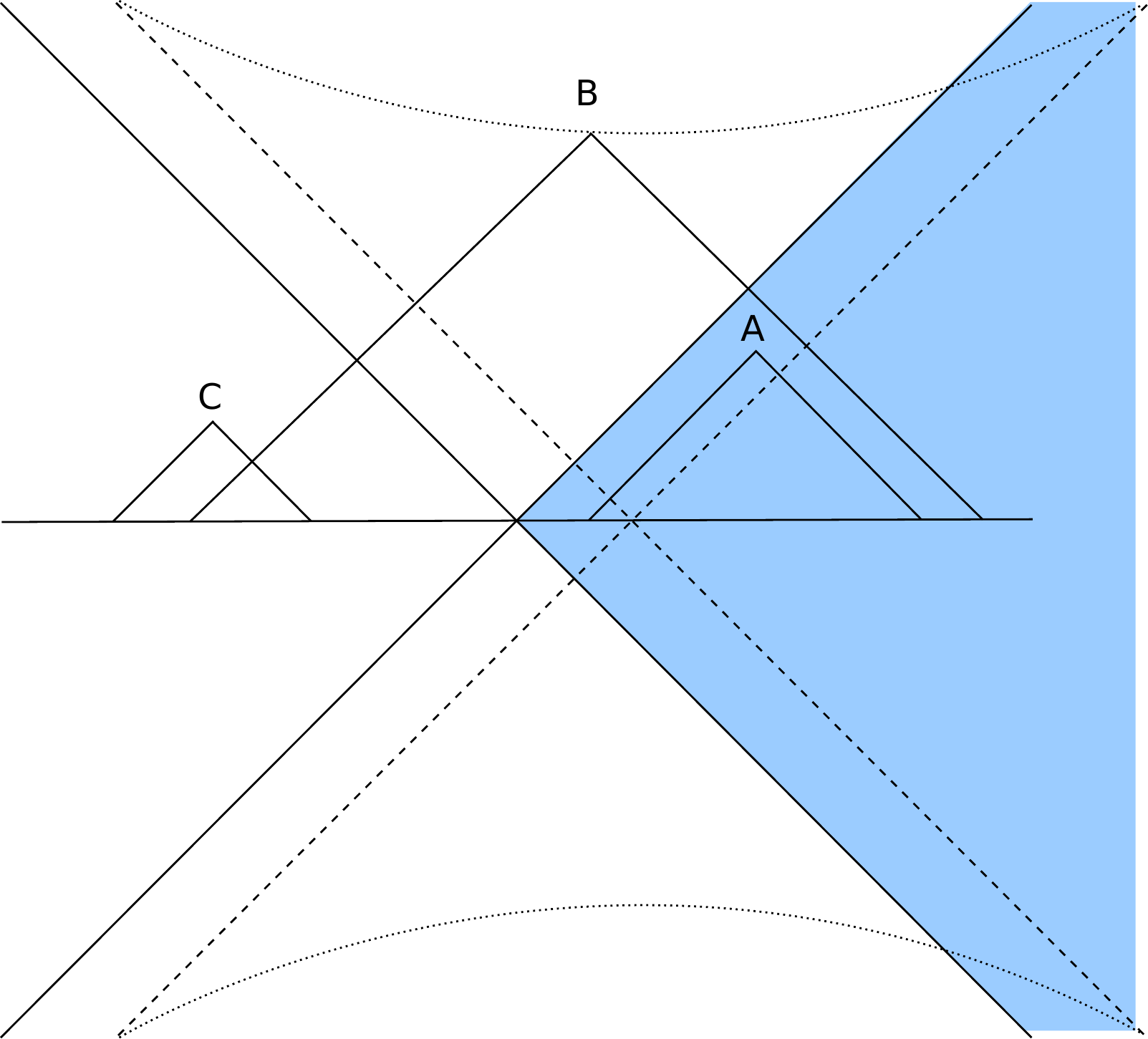}
\caption{Left: $t=0$ Cauchy slice, with a bulk point $x$ displaced slightly into the eastern Rindler wedge (shaded). Right: time-evolution of local boundary operators to bilocals at $t=0$. The dashed axes show the light-cone of the bulk point; the future and past singularities in the smearing function are indicated by the dotted lines. Point $A$ falls entirely within the eastern wedge, while one leg of $B$, and both legs of $C$, must be mapped into the east using using the entanglement of the Rindler vacuum. Note that with the bulk point as shown, at most one singular leg  must be mapped, but this potential divergence is exactly cancelled by a decaying exponential arising from \eqref{eq:MinkEnt}, so the resulting expression remains well-defined.\label{fig:wedgeconnected}}
\end{figure}

From the expression \eqref{eq:Laurensbilocal}, one sees that upon time-evolving the boundary operator $\mathcal{O}=\pd_+\phi\pd_-\phi$ to the $t=0$ Cauchy slice, one or both parts of the resulting bilocal may have support in the western wedge (indicated by $\beta^W$). We now demonstrate that the entanglement present in the Minkowski vacuum can be used to map these parts into the east. The set-up is illustrated schematically in fig. \ref{fig:wedgeconnected}. 

The key observation is that acting on the Minkowski vacuum with a Rindler operator, we have
\eq{
\beta^W_{\omega_\pm}|0\rangle=e^{-\pi \omega_\pm} \beta^E_{-\omega_\pm}|0\rangle\label{eq:MinkEnt}
}
which one can see by writing $|0\rangle\propto\bigotimes_\omega\sum_n e^{-\pi \omega n} |n\rangle_W \otimes |n\rangle_E$. We shall use this fact to write \eqref{eq:Laurensbilocal} entirely in terms of operators in the eastern wedge, $\beta^E$. For this mapping between western and eastern operators  to succeed, we require only that both their left- and right-action on the vacuum state agree,
\be
\Phi \ket{0}=\mathcal{O}_E\ket{0}~,\;\;\;
\mathrm{and}\;\;\;
\bra{0} \Phi=\bra{0}\mathcal{O}_E~,\label{eq:mapcondition}
\ee
which is enough to ensure that 2-pt correlators are preserved. Our strategy is to satisfy the left equation by construction, and then check whether the right equation is also satisfied. 

Performing this mapping allows us to write \eqref{eq:Laurensbilocal} as
\ben
\ba
\Phi(0,a,z)|0\rangle=\iint\dd\omega_+\dd\omega_-&\left(-\tilde K_N\lp\omega_+,\omega_-\rp \beta^E_{\omega_+}\beta^E_{-\omega_-}e^{-\pi \omega_-} - \tilde K_S\lp-\omega_+,-\omega_-\rp\beta^E_{-\omega_+}\beta^E_{\omega_-}e^{-\pi \omega_+}\right .\\
&\left.+ \tilde K_E\lp\omega_+,-\omega_-\rp \beta^E_{\omega_+}\beta^E_{\omega_-}+ \tilde K_W\lp-\omega_+,\omega_-\rp\beta^E_{-\omega_+}\beta^E_{-\omega_-}e^{-\pi(\omega_+ + \omega_-)}\right)|0\rangle\\
=\iint\dd\omega_+\dd\omega_-&\left(-\cosh(\pi\omega_+) e^{-\pi \omega_-} - \cosh(\pi\omega_-)e^{\pi \omega_+}\right.\\
&\left.+\cosh(\pi(\omega_+-\omega_-))+ e^{\pi(\omega_+ - \omega_-)}\right)\tilde K_W\lp\omega_+,\omega_-\rp\beta^E_{\omega_+}\beta^E_{-\omega_-}|0\rangle\\
=-2\iint\dd\omega_+\dd\omega_-&\sinh(\pi\omega_+)\sinh(\pi \omega_-) \tilde K_W\lp\omega_+,\omega_-\rp \beta^E_{\omega_+}\beta^E_{-\omega_-}|0\rangle
\ea
\een
where have used the relations \eqref{eq:Ben}. Substituting in the explicit form of $\tilde K_W$, \eqref{eq:fourthterm},
we find
\eq{
\Phi|0\rangle=-2\pi^2\lp\frac{z}{a}\rp^2\int\dd\omega_+\dd\omega_-a^{-i(\omega_+ + \omega_-)}{}_2F_1\lp1+i \omega_+,1+i\omega_-,2,-\frac{z^2}{a^2}\rp\beta^E_{\omega_+}\beta^E_{-\omega_-} |0\rangle~.\label{eq:PhiEnt}
}
which is precisely \eqref{eq:PhiK}, with $\tilde{\mathcal{O}}^E_{\omega_+,\omega_-}=\beta^E_{\omega_+}\beta^E_{-\omega_-}$. One can check that this operator $\Phi$ satisfies the condition \eqref{eq:mapcondition}. This demonstrates that the entanglement structure of Minkowski space can be used to localize the precursor entirely within a single Rindler wedge, thus providing an alternative realization of the approach based on gauge freedom discussed above.

\subsection{Mapping the precursor into the ``wrong'' Rindler wedge}
To further explore this link between precursors and entanglement, let us now ask what happens if we instead attempt to map the bilocal operator into the western wedge. Since our bulk point is located in the east, we would na\"ively expect this to fail, as this would correspond to reconstructing the bulk operator located outside the causal/entanglement wedge (cf. the end of section \ref{sec:dK}). Hence we refer to this as mapping the precursor into the \emph{wrong} wedge.

\begin{figure}[h!]
\centering
\includegraphics[width=0.35\textwidth]{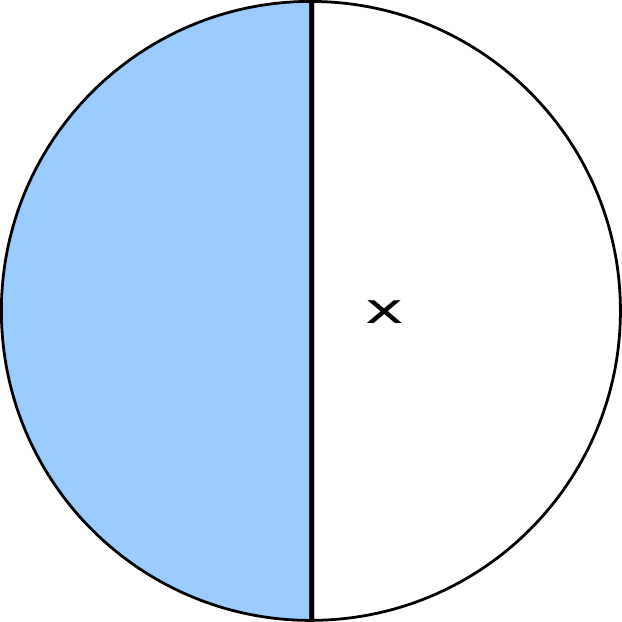}
\hspace{1 cm}
\includegraphics[width=0.4\textwidth]{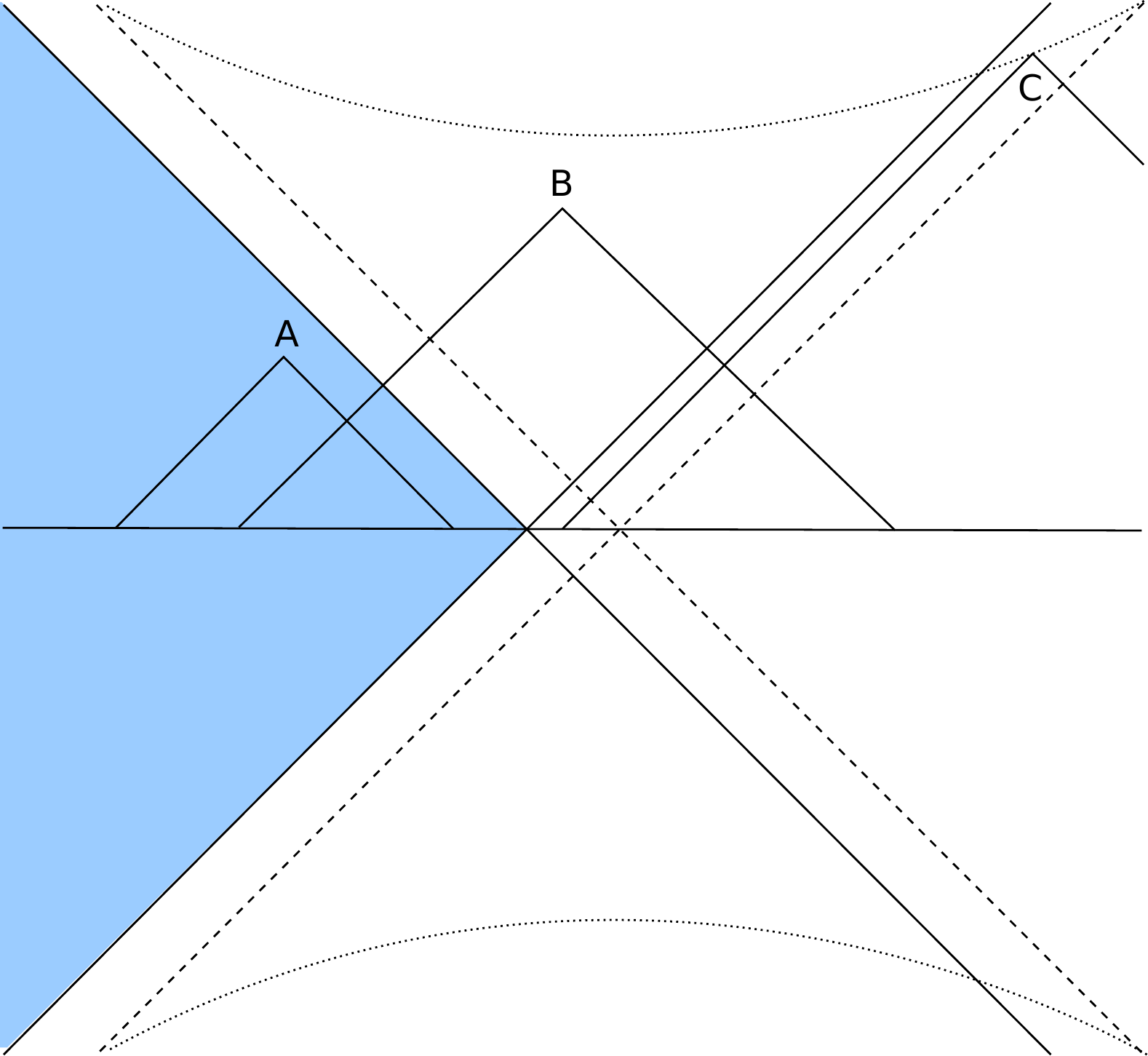}
\caption{Left: $t=0$ Cauchy slice, with a bulk point $x$ displaced slightly into east as before, but reconstruction attempted in the western (wrong) wedge Rindler wedge (shaded). Right: time-evolution of local boundary operators to bilocals at $t=0$. Note that while $A$ and $B$ can be mapped without difficulty, as discussed in the previous section, there are now points like $C$ with two divergent legs, both of which must be mapped into the western wedge. This is one more exponential in momentum than we are capable of taming, and thus localization of the associated bulk point fails.\label{fig:wrong}}
\end{figure}

The set-up is illustrated in fig. \ref{fig:wrong}. Following the same procedure as in the previous subsection, one obtains
\be
\ba
\Phi(0,a,z) |0\rangle &=-2\int\dd\omega_+\dd\omega_- e^{\pi(\omega_- -\omega_+)} \sinh(\pi\omega_+)\sinh(\pi \omega_-) \tilde K_W\lp\omega_+,\omega_-\rp \beta^W_{-\omega_+}\beta^W_{\omega_-}|0\rangle\\
&\propto \int\dd\omega_+\dd\omega_- a^{-i(\omega_- +\omega_+)}e^{\pi(\omega_- -\omega_+)}{}_2F_1\lp1+i \omega_+,1+i\omega_-,2,-\frac{z^2}{x^2}\rp\beta^W_{-\omega_+}\beta^W_{\omega_-} |0\rangle ~.
\label{eq:Phiwrong}
\ea
\ee

But upon conjugating \eqref{eq:Phiwrong}, and taking $\omega_\pm\rightarrow-\omega_\pm$ under the integral, we find
\eq{
\Phi^\dagger\ket{0}\propto \int\dd\omega_+\dd\omega_- a^{-i(\omega_- +\omega_+)}e^{-\pi(\omega_- -\omega_+)}{}_2F_1(1+i \omega_+,1+i\omega_-,2,-\frac{z^2}{x^2}) \beta^W_{-\omega_+}\beta^W_{\omega_-}\ket{0}~,
}
in clear violation of the condition \eqref{eq:mapcondition}. Thus our entanglement mapping condition fails when the bulk operator lies in the complement of the selected boundary region. 

The importance of this condition was recently emphasized in \cite{Dong_Harlow_Wall_2016}, who phrased it as the requirement of hermiticity. In particular, they proved that in order to satisfy \eqref{eq:mapcondition}, the field $\Phi$ must lie within the bulk entanglement wedge of the boundary region that contains the operator $\mathcal{O}$. Our model may therefore be taken as an explicit demonstration of this principle. Specifically, if one attempts to localize the boundary representation of a bulk operator in the complement, the resulting operator will be non-hermitian. In order to construct a well-defined precursor, the localization must be attempted within the entanglement wedge that includes the bulk field in question. 

One can see that the wrong-wedge operator \eqref{eq:Phiwrong} is manifestly ill-behaved when acting on the Minkowski vacuum: in the limit $\omega_+ \gg 1$ and $\omega_- \ll -1$, we have two Rindler creation operators acting on $|0\rangle$, with a coefficient which grows exponentially. This means we create a state which is highly UV-sensitive (note the singular legs that must be mapped in fig. \ref{fig:wrong}). Indeed, one can show that the two-point function $\langle 0 | \Phi \Phi^\dagger |0\rangle$ diverges using the wrong-wedge operator $\Phi$. The fact that UV-divergences occur in the same circumstance as when hermiticity is lost is suggestive, but we have not found a clear conceptual link between the two.

\section{Discussion}\label{sec:discussion}
The boundary duals of operators deep in the bulk have highly non-local representations in the CFT, known as ``precursors''. Following the HKLL construction, these can be localized to the boundary region of an AdS-Rindler wedge that contains the bulk field. \cite{HKLL_2006} This immediately raises the question of redundant boundary duals: as illustrated in \cite{ADH}, a bulk field that falls within multiple boundary wedges must have multiple, different boundary representations. We are therefore left with the problem of how inequivalent precursors can all give rise to the same bulk operator.

In this paper, we presented an improved version of the model in \cite{MPR}, wherein it was argued that the non-uniqueness of precursors is a simple consequence of boundary gauge invariance. We have provided an explicit demonstration of this proposal, using the gauge freedom in the smearing function to localize precursors within a single Rindler wedge. This supports the claim \cite{MPR} that gauge invariance may be deeply connected to the emergence of the dual spacetime. In section \ref{sec:dK}, this was accomplished without any mention of boundary entanglement. Rather, it relied only on the freedom to add pure-gauge modes to the precursor.

In contrast, entanglement is essential for a quantum error correction scheme to succeed \cite{ADH}. Indeed, it has been postulated that the entanglement between boundary regions plays a crucial role in the emergence of the bulk spacetime, and there are reasons to believe that the entanglement -- as opposed to the causal -- wedge is the more natural bulk dual for holographic reconstruction \cite{gravdual, LM_2013,Dong_Harlow_Wall_2016,FLM_2013,Headrick_etal_2014,Bao_Kim_2016,Wall_2012}. In the interest of further exploring the link between entanglement and localization, we showed explicitly in section \ref{sec:entanglement} that the entanglement between boundary Rindler wedges can likewise be used to localize information to within a single region, in agreement with the approach from gauge freedom above.

To some extent, the distinction between gauge invariance and error correction is linguistic. In writing the pure-gauge operators in a particular form, in terms of creation and annihilation operators, MPR \cite{MPR} make use of the large-$N$ approximation, which includes an assumption about the class of states. So although the freedom to add these operators was demonstrated by MPR to be linked to gauge invariance, their explicit form, and thus the resulting localized precursor, does rely on an assumption that we act within the low-energy subspace of the theory.\footnote{In fact, this issue arises already at the level of the smearing function. As discussed in \cite{Leich_Rosen_2013,Bousso_etal_2013,Morrison_2014}, there are subtleties in attempting to construct an HKLL-type precursor in non-trivial geometries, e.g. in the presence of horizons.} We hope that we have shed some light on the physics, if not the linguistics, by explicitly calculating the resulting operators.

Our model may also be useful for diagnosing proposals for the description of operators behind the black hole horizon, such as \cite{Guica_Jafferis_2015}, since the bulk spacetime we considered does have a Rindler horizon. In addition, it may clarify subtleties in the CFT operators dual to bulk fields outside the black hole horizon, which have the same properties as our Rindler precursors.

\begin{figure}[h!]
\centering
\includegraphics[width=0.35\textwidth]{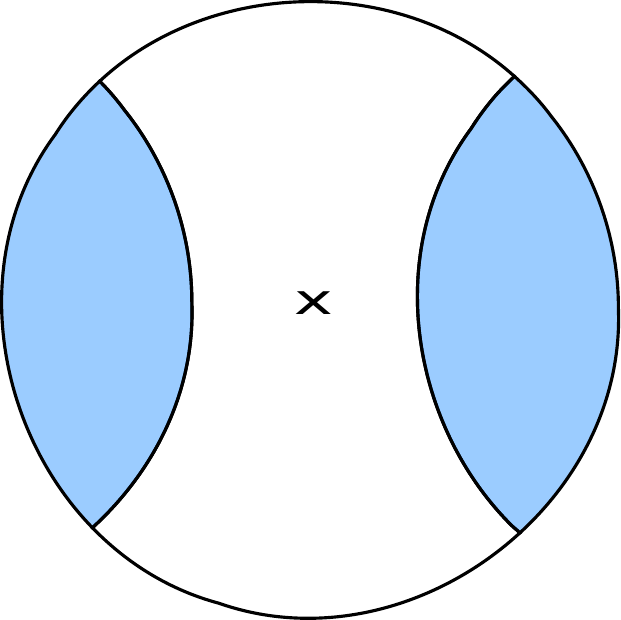}
\hspace{1 cm}
\includegraphics[width=0.35\textwidth]{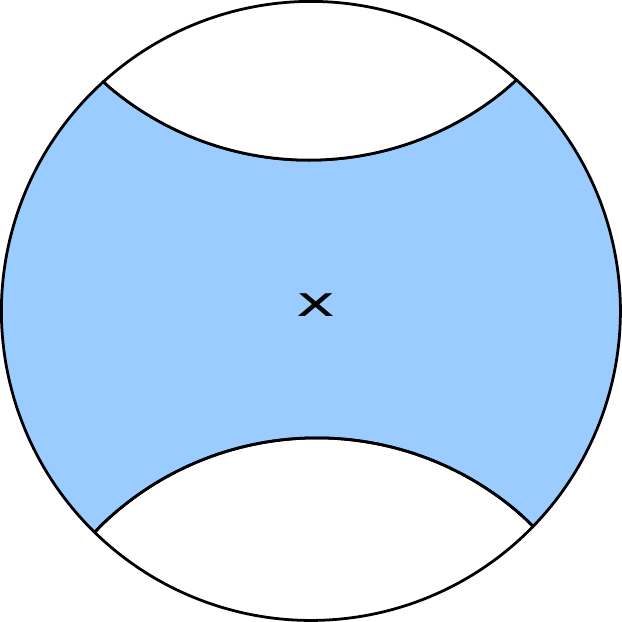}
\caption{Entanglement wedge for a disconnected region, shaded. If the region is sufficiently small (left), the bulk point labelled $x$ will not be included, and hence the shaded boundary region contains no information about it. However, as the boundary region is increased, the bulk geodesics that define the entanglement wedge eventually transition to a new global minimum (right), whereupon the shaded boundary region abruptly gains information about the given bulk point. Intuitively, one needs ``enough'' of the boundary to reconstruct the bulk.\label{fig:wedgedisconnected}}
\end{figure}

It is interesting to ask whether our model continues to agree with expectations about the full AdS/CFT correspondence when we consider more complicated boundary regions, such as disconnected intervals. The analogous set-up for a disconnected boundary region is shown in figure \ref{fig:wedgedisconnected}. The shaded region is the entanglement wedge for the given, disconnected boundary region. When this region becomes sufficiently large, the bulk minimal surface transitions to the new global minimum, whereupon the entanglement wedge suddenly includes the bulk point \cite{Hubeny_etal_2013,Freivogel_etal_2014}. The question we wish to ask is whether our model generalizes to agree with the corresponding reconstruction prescription. Specifically, can the precursor corresponding to a bulk point within the shaded bulk region be localized within the (disconnected) 
boundary of this region?

Generalizing our above results to multiple, disconnected boundary regions requires either an explicit formula for the pure-gauge smearing function $\delta K$, or a general prescription for when a particular bilocal can be mapped into a given wedge. We do not present a general solution here, but instead comment on what one might expect given the above results, in the interest of comparing them with reconstruction proposals involving the entanglement wedge \cite{Dong_Harlow_Wall_2016} and quantum error correction \cite{ADH}. 

\begin{figure}[h!]
\centering
\includegraphics[width=0.95\textwidth]{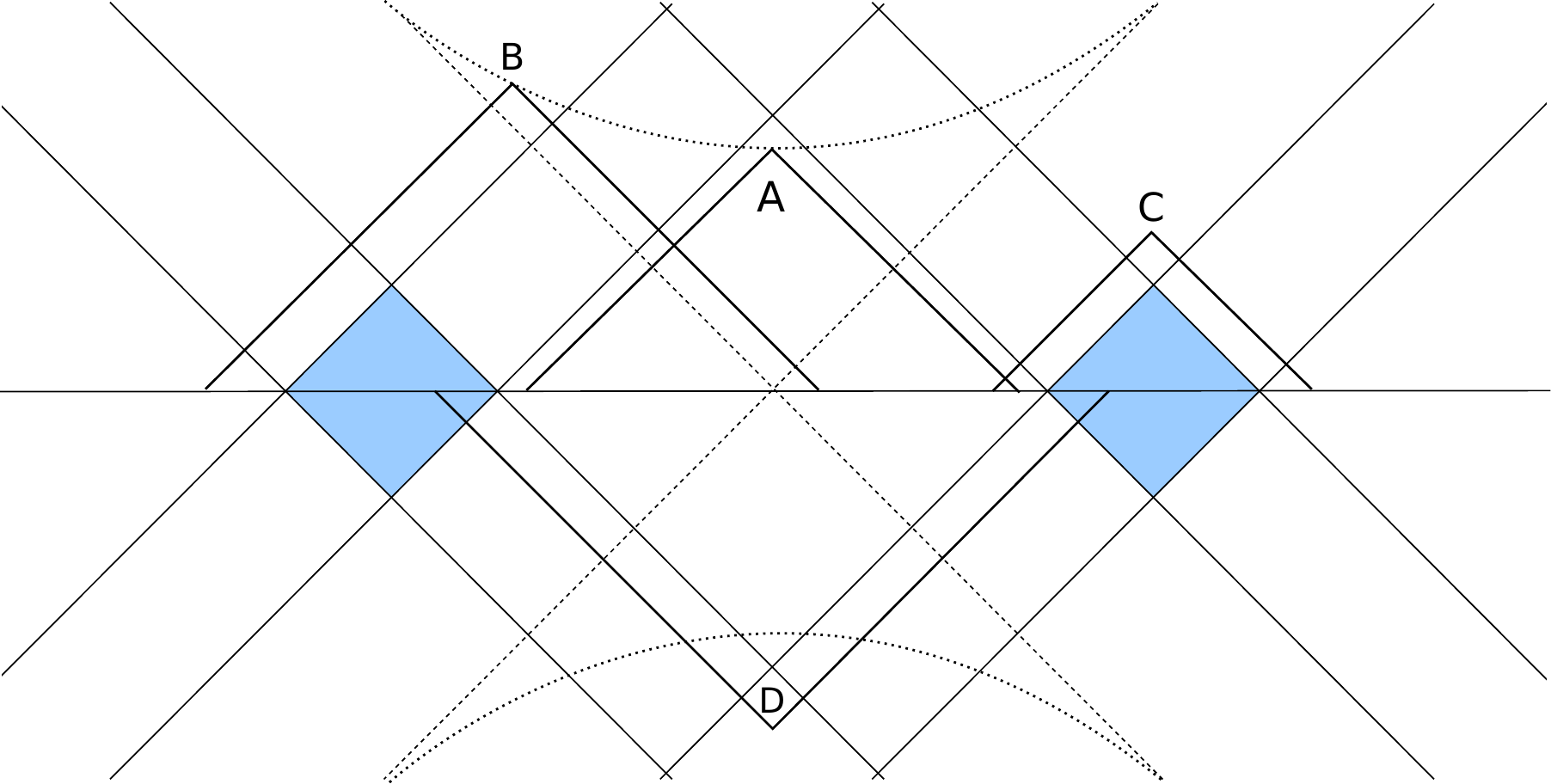}
\caption{Disconnected region shown in fig. \ref{fig:wedgedisconnected}, with the bulk point in the center. Points $A$ and $B$ each have two singular legs, and cannot be mapped into the correct (shaded) boundary region with the na\"ive extension of our model. Which of $C$ or $D$ requires mapping is highly model-specific. In our model, with information localized along the edge of the light cone, only $C$ requires mapping. If one instead devises a model in which information is smeared out along spacelike slices of the light cone, then most of $C$ would instead fall into the correct wedge, but $D$ would require a (presumably more complicated) mapping procedure.\label{fig:multipatch}}
\end{figure}

Figure \ref{fig:multipatch} demonstrates the potential problem. Na\"ively generalizing our results above for the case of a single boundary region, we suspect that bilocals with both points outside the boundary region, which are in addition integrated against non-smooth functions, cannot be mapped to healthy operators within the given CFT region. The smearing function K is smooth except at the boundary points that are lightlike connected to the bulk point, indicated by the dotted hyperbolas in the figure. Our suggested criterion, then, is that the precursor cannot be localized within the boundary wedge if some bilocals that are evolved back from the lightcone singularity have both points outside our region of the CFT.

Referring to the figure, one can see that even a bulk field within the entanglement wedge (the right image in \ref{fig:wedgedisconnected}) leads to such divergent bilocals that we cannot map into the correct boundary region. These are indicated by points $A$ and $B$ in fig. \ref{fig:multipatch}. Therefore, if our guess is correct for when the precursor can be localized, our simple model fails to reproduce the expected result, namely that bulk operators in the entanglement wedge can be mapped to precursors in the corresponding boundary region.

This should perhaps not be too surprising, since expectations about the entanglement wedge are based on the Ryu-Takayanagi formula for the entanglement entropy. However, it is known that a simple free field model on the boundary will not reproduce the correct RT formula for the entanglement entropy of multiple intervals after a quench \cite{Leichenauer_Moosa_2015}. So it may simply be that our weakly-coupled model does not preserve the requisite entanglement between subregions upon evolving to bilocals along a single Cauchy slice. 

One is therefore led to ask whether our model can be improved to capture the entire bulk entanglement wedge. Consider point $C$ in fig. \ref{fig:multipatch}. In our model, this is time-evolved to bilocals lying entirely outside the entanglement wedge. However, one could imagine a different model in which the information about the local operator at $C$ becomes smeared out along the intersection of the backwards lightcone of $C$ and the $t=0$ Cauchy slice, such that this point is still captured -- that is, requires no potentially problematic mapping -- provided some minimum amount of information falls within the wedge, perhaps evoking some quantum secret sharing scheme \cite{QSS}.

Unfortunately, such a modification creates other problems. Consider instead point $D$, which is time-evolved to bilocals lying entirely within the disconnected wedge. If we instead adopt this modification to our model, then most of the information about $D$ would fall outside this region. Thus, in terms of mapping difficulty, we've only succeeded in trading $C$ for $D$, and the underlying problem remains.

Performing the localization via a pure-gauge smearing function $\delta K$ likewise appears impossible, although it would be interesting to try to extend our techniques to this case. Recall from section \ref{sec:dK} that the ability to fix $\hat K=K+\delta K=0$ relied on nontrivial relationships between the smearing function in different wedges in order to reduce the number of conditions (one per wedge) from three down to two, the number of undetermined coefficients. While we have not proven it, a quick glance at the many wedges of fig. \ref{fig:multipatch} -- which contains four Rindler-type axes -- suggests that fixing $\hat K$ to zero in all but the two shaded regions would require a miraculous conspiracy of conditions.

Thus, while our model appears to generalize naturally to disconnected \emph{causal} wedges, there is no obvious generalization that would correctly reproduce the \emph{entanglement} wedge prescription. However, one fully expects that in the latter case a localized operator satisfying the condition \eqref{eq:mapcondition} exists. Understanding precisely how the entanglement structure, or the gauge freedom, conspires to produce localized precursors for more general boundary regions would be illuminating. 

\section*{Acknowledgements}
We thank Davide Dispenza, Don Marolf, Ian Morrison, Joe Polchinski, and Vladimir Rosenhaus for helpful discussions. We would also like to thank the Kavli Institute for Theoretical Physics for their hospitality during the \emph{Quantum Gravity Foundations: UV to IR} program, where this work was initiated. Support at KITP was provided in part by the National Science Foundation under Grant No. NSF PHY11-25915. This work is part of the $\Delta$-ITP consortium and supported in part by the Foundation for Fundamental Research on Matter (FOM); both are parts of the Netherlands Organization for Scientific Research (NWO) funded by the Dutch Ministry of Education, Culture, and Science (OCW). 

\begin{appendix}
\section{Relating Fourier transforms of the smearing function}\label{app:fourier}
In this appendix, we will prove the relations \eqref{eq:Ben}:
\be
\ba
\tilde K_N\lp\omega_+,\omega_-\rp&=\cosh(\pi\omega_+)\tilde K_W\lp\omega_+,\omega_-\rp\\
\tilde K_S\lp\omega_+,\omega_-\rp&=\cosh(\pi\omega_-)\tilde K_W\lp\omega_+,\omega_-\rp \label{eq:Benapp}\\
\tilde K_E\lp\omega_+,\omega_-\rp&=\cosh\lp\pi(\omega_+-\omega_-)\rp \tilde K_W\lp\omega_+,\omega_-\rp
\ea
\ee
where the Fourier transform of the smearing function, $\tilde K_W$, is given by \eqref{eq:FourierK}, with $K$ written in light-cone coordinates as in \eqref{eq:poincareKLC}:
\be
\ba
\tilde K_W\lp\omega_+,\omega_-\rp&= \int_{-\infty}^0\dd x_+\int_0^{\infty}\dd x_- \log\left(\frac{|z^2-(x_+-a)(x_- +a)|}{2z} \right)|x_+|^{-i \omega_+-1}|x_-|^{-i \omega_- -1} \\
&= \int_0^\infty\dd x_+\int_0^{\infty} \dd x_- \log\left(\frac{|z^2+(x_++a)(x_- +a)|}{2z} \right)|x_+|^{-i \omega_+-1}|x_-|^{-i \omega_- -1} \label{eq:Wfourierintegral} \\
&= \int_{-\infty}^\infty\dd u \int_{-\infty}^{\infty}\dd v \log\left(\frac{|z^2+(e^u+a)(e^v +a)|}{2z} \right)e^{-i \omega_+ u}e^{-i \omega_- v}~,
\ea
\ee
where in the last step we've made the change of variables $x_+=e^u$, $x_-=e^v$. Note that the logarithm does not become singular in the western quadrant, as we shifted the bulk point into the east (cf. fig. \ref{fig:wedgeconnected}). For convenience, we may rescale the zero mode to remove the constant factor in the denominator of the argument of the logarithm. Hence, suppressing the $\omega_+$ and $\omega_-$ subscripts, the explicit expression for $\delta K$ in each of the four wedges may be written
\eq{
\tilde K_W= \int_{-\infty}^\infty \dd u \int_{-\infty}^{\infty}\dd v \log\left(|{z^2+(e^u+a)(e^v +a)|}{} \right)e^{-i \omega_+ u}e^{-i \omega_- v}\label{eq:Wapp}
}
\eq{
\tilde K_N= \int_{-\infty}^\infty\dd u \int_{-\infty}^{\infty}\dd v \log\left(|{z^2-(e^u-a)(e^v +a)|}{} \right)e^{-i \omega_+ u}e^{-i \omega_- v} \label{eq:Napp}
}
\eq{
\tilde K_E= \int_{-\infty}^\infty\dd u \int_{-\infty}^{\infty}\dd v \log\left(|{z^2+(e^u-a)(e^v-a )|}{} \right)e^{-i \omega_+ u}e^{-i \omega_- v} \label{eq:Eapp}
}
\eq{
\tilde K_S= \int_{-\infty}^\infty\dd u \int_{-\infty}^{\infty}\dd v \log\left(|{z^2-(e^u+a)(e^v -a)|}{} \right)e^{-i \omega_+ u}e^{-i \omega_- v} .
}

Let us begin by relating $\tilde K_W$ and $\tilde K_N$. Define the function $f(u)$ for $u\in\mathbb{C}$ as\footnote{For simplicity we included the $v$-integral in the definition of $f$. For the reader worried about its convergence, the following contour argument can still be made, relating the $v$-integrands, by defining $f(u)\equiv e^{-i \omega_- v} \log\left({z^2+(e^u+a)(e^v +a)}{} \right)e^{-i \omega_+ u}$ for fixed $v$.}
\eq{
f(u)\equiv \int_{-\infty}^\infty\dd v \log\left({z^2+(e^u+a)(e^v +a)}{} \right)e^{-i \omega_+ u}e^{-i \omega_- v}~.\label{eq:fdef}
}
Note that integrating $f$ over the real $u$-axis gives $\tilde K_W$ (since $a>0$), while integrating $f(u\pm i \pi)$ is of the same basic form as $\tilde K_N$,
\eq{
f(u\pm i \pi)=e^{\pm \pi \omega_+}\int_{-\infty}^\infty\dd v \log\left(z^2-(e^{{u}}-a)(e^v +a) \right)e^{-i \omega_+ {u} }e^{-i\omega_- v}~,\label{eq:f2def}
}
up to a factor of $e^{\pm\pi\omega_+}$, and ambiguities due to the singularities in the logarithm. In particular, the argument of the log is negative when $z^2<\lp e^u-a\rp\lp e^v+a\rp$, so there is a branch point at
\be
u^*=\log\lp\frac{z^2}{e^v+a}+a\rp
\ee
and branch cuts running horizontally at $u\pm i \pi$ for $u>u^*$. Now, imagine a rectangular contour in the complex $u$-plane running from $-\infty$ to $\infty$ along the real axis, and then back the other way along $u\pm i\pi$ (that is, just inside the complex region prescribed by the branch cuts). Since we enclose no poles, the total contour integral vanishes, and we may write
\be
0=\int_{-\infty}^\infty\dd u\lp f(u)-f(u\pm i\pi)\rp
+\int_{\infty}^{\infty\pm i\pi}\dd uf(u)
+\int_{-\infty\pm i\pi}^{-\infty}\dd uf(u)\label{eq:contourexplain}
\ee
where the last two terms are the vertical side contributions for the function evaluated at $u\rightarrow\pm\infty$ from the real axis to $\pm i\pi$. One takes the upper signs in \eqref{eq:contourexplain} for the contour in the upper half-plane, which runs counter-clockwise, and the lower signs for the clockwise contour in the lower half. We then observe that the side contributions can be made to vanish by suitably deforming the contour off the real axis. Hence, dropping these terms and writing the above expression in terms of $\tilde K$, we have
\eq{
0=\tilde K_W-e^{\pm\pi\omega_+}\tilde K_N-e^{\pm\pi\omega_+}\int_{u^*}^\infty\dd u \int_{-\infty}^\infty\dd v\lp \pm i\pi\rp e^{-i \omega_+ u}e^{-i \omega_- v}\label{eq:relationpre}
}
where we've taken the principle value of the complex logarithm in \eqref{eq:f2def}, $\mathrm{Log}(x)=\log|x|\pm i\pi$, where the upper/lower sign corresponds to approaching the negative real axis from the upper/lower half-plane, respectively (i.e., our choice of contour). By a linear combination of the two equations in \eqref{eq:relationpre}, the third term on the r.h.s. cancels, and one obtains
\be
\tilde K_N=\cosh(\pi \omega_+)\tilde K_W
\ee
which is the desired result. Similarly, one can show
\eq{\tilde K_S=\cosh (\pi \omega_-) \tilde K_W~.}

The derivation of the third relation, between $\tilde K_W$ and $\tilde K_E$, follows a similar contour argument, but requires a slight change of coordinates. In particular, we first write \eqref{eq:Eapp} as
\eq{
\tilde K_E= \int_{-\infty}^\infty\dd t \int_{-\infty}^{\infty} \dd x \log\left(|{z^2+(e^{t+x}-a)(e^{t-x}-a )|}{} \right)e^{-i \omega t}e^{-i k x}\label{eq:Erewrite}
}
where we defined $\omega\equiv \omega_+ + \omega_-$ and $k\equiv \omega_- - \omega_+$, and similarly for $\tilde K_W$. We then define a function $g(x)$ for $x \in \mathbb{C}$,
\eq{
g(x)\equiv \int_{-\infty}^\infty\dd t \log\left({z^2+(e^{t+x}-a)(e^{t-x}-a )}{} \right)e^{-i \omega t}e^{-i k x} ~,
}
which will be related to \eqref{eq:Erewrite} upon integrating along the $x$-axis, and observe that the integral of
\eq{g(x\pm i \pi)\equiv \int_{-\infty}^\infty\dd t \log\left({z^2+(e^{t+x}+a)(e^{t-x}+a )}{} \right)e^{-i \omega t}e^{-i k x} e^{\pm \pi k}}
yields $\tilde K_W$. 

We can now apply essentially the same argument as before. The argument of the logarithm in \eqref{eq:Erewrite} is negative when $z^2+a^2+e^{2t}< 2ae^t\cosh x$, implying branch points at
\be
x^*=\pm\cosh^{-1}\lp\frac{z^2+a^2+e^{2t}}{2ae^t}\rp~.
\ee
We choose the branch cuts running out horizontally to infinity. The integration contours are then restricted to the rectangular region between the $x$-axis and $x\pm i\pi$, given an expression analogous to \eqref{eq:contourexplain}. Analytically continuing the logarithm to complex values as above, and dropping the side contributions, we have
\eq{
0=\tilde K_E-e^{\pm \pi k }\tilde K_W +\lp\int_{-\infty}^{-x^*}\dd x +\int_{x^*}^\infty\dd x\rp\int_{-\infty}^\infty\dd t(\pm i\pi)e^{-i \omega t}e^{-i k x}~.
}
Taking a linear combination of these two equations, we obtain
\eq{
\tilde K_E=\cosh(\pi(\omega_- - \omega_+))\tilde K_W~,
}
as desired.

\section{Evaluating the smearing function}\label{app:integral}
In this appendix we evaluate the Fourier integral of the smearing function in the western Rindler wedge, $\tilde K_W$ \eqref{eq:Wapp},
\be
\tilde K_W= \int_{-\infty}^\infty \dd u \int_{-\infty}^{\infty}\dd v \log\left({z^2+(e^u+a)(e^v +a)}{} \right)e^{-i \omega_+ u}e^{-i \omega_- v}~,
\ee
where the argument of the log is always positive by virtue of our having shifted the bulk point into the east, as described in the main text. Integrating by parts twice, this becomes
\ben
\ba
\tilde K_W=
-&\frac{1}{\omega_+\omega_-}e^{-i\omega_+u}e^{-i\omega_-v}\ln\lp z^2+\lp e^u+a\rp\lp e^v+a\rp\rp\bigg|_{u, v=-\infty}^\infty\\
+&\frac{1}{\omega_+\omega_-}\int_{-\infty}^\infty\dd ue^{-i\omega_+u}e^{-i\omega_-v}\frac{e^u\lp e^v+a\rp}{z^2+\lp e^u+a\rp\lp e^v+a\rp}\bigg|_{v=-\infty}^\infty\\
+&\frac{1}{\omega_+\omega_-}\int_{-\infty}^\infty\dd ve^{-i\omega_+u}e^{-i\omega_-v}\frac{\lp e^u+a\rp e^v}{z^2+\lp e^u+a\rp\lp e^v+a\rp}\bigg|_{u=-\infty}^\infty\\
-&\frac{z^2}{\omega_+\omega_-}\int_{-\infty}^\infty\dd u\dd ve^{-i\omega_+u}e^{-i\omega_-v}\frac{e^{u+v}}{\lp z^2+\lp e^u+a\rp\lp e^v+a\rp\rp^2}~.
\ea
\een
The first three (boundary) terms can be made to vanish by a suitable contour deformation. The remaining double integral (the fourth term) can be evaluated to yield
\be
\tilde K_W=
-\pi^2\lp\frac{z}{a}\rp^2a^{-i\lp\omega_++\omega_-\rp}\csch(\pi\omega_+)\csch(\pi\omega_-){}_2F_1\lp1+i\omega_+,1+i\omega_-,2,\frac{-z^2}{a^2}\rp~.
\label{eq:fourthterm}
\ee

\section{Computing the two-point function}
\label{app:twopoint}

As an extra check of our formalism, we include an explicit calculation of the two-point function, and show that it reduces to the correct $\mathrm{AdS}_{2+1}$ correlator in the near-horizon limit. This will serve as a diagnostic of whether our expression for the bulk field constructed from boundary data entirely in the eastern wedge, \eqref{eq:PhiEnt}
\ben
\Phi(0,a,z)=-2\pi^2\lp\frac{z}{a}\rp^2\int\dd\omega_+\dd\omega_-a^{-i(\omega_++\omega_-)}{}_2F_1\lp1+i\omega_+,1+i\omega_-,2,\frac{-z^2}{a^2}\rp\beta^E_{\omega_+}\beta^E_{-\omega_-}~,
\een
is well-defined. Here $\beta^E_{\omega_\pm}$ are the Rindler creation ($\omega<0$) and annihilation ($\omega>0$) operators in the eastern wedge, as defined in the main text. Since we work entirely in the eastern wedge in what follows, we shall henceforth suppress the superscript $E$ to minimize clutter.

Inside the two-point function, we will have left/right moving Rindler operators acting on the Minkowski vacuum. As the left- and right-movers commute, the four-$\beta$ correlator is
\eq{
\langle 0|\beta_{\omega_+}\beta_{-\omega_-}\beta_{\omega'_+}\beta_{-\omega'_-}|0\rangle=\delta(\omega_+ + \omega'_+)\delta(\omega_- + \omega'_-) \left(\frac{\omega_+}{1-e^{-2\pi \omega_+}} \right)\left(\frac{\omega_-}{e^{2\pi \omega_- }-1}\right)~.
}
The bulk two-point function we seek to examine is therefore written explicitly as
\ben
\ba
\left<\Phi(a_1,z_1)\Phi(a_2,z_2)\right>=&4\pi^4\lp\frac{z_1z_2}{a_1a_2}\rp^2
\int_{-\infty}^\infty\dd\omega_+\dd\omega_-\dd\omega_+'\dd\omega_-'\delta(\omega_++\omega_+')\delta(\omega_-+\omega_-')\\
\times&a_1^{-i\lp \omega_++\omega_-\rp}a_2^{-i\lp \omega_+'+\omega_-'\rp}
\lp\frac{\omega_+}{1-e^{-2\pi\omega_+}}\rp\lp\frac{-\omega_-}{1-e^{2\pi\omega_-}}\rp\\
\times&{}_2F_1\lp1+i\omega_+,1+i\omega_-,2,\frac{-z_1^2}{a_1^2}\rp{}_2F_1\lp1+i\omega_+',1+i\omega_-',2,\frac{-z_2^2}{a_2^2}\rp~.
\ea
\een
By virtue of the delta functions, the integrals over primed frequencies are trivial:
\be
\ba
\left<\Phi(a_1,z_1)\Phi(a_2,z_2)\right>=&\pi^4\lp\frac{z_1z_2}{a_1a_2}\rp^2
\int_{-\infty}^\infty\dd\omega_+\dd\omega_-\lp\frac{a_1}{a_2}\rp^{-i\lp\omega_++\omega_-\rp}\\
\times&\omega_+\omega_-\lp\coth(\pi\omega_+)+1\rp\lp\coth(\pi\omega_-)-1\rp\\
\times&{}_2F_1\lp1+i\omega_+,1+i\omega_-,2,\frac{-z_1^2}{a_1^2}\rp{}_2F_1\lp1-i\omega_+,1-i\omega_-,2,\frac{-z_2^2}{a_2^2}\rp~.\label{eq:2pttemp}
\ea
\ee
Unfortunately, we have not succeeded in evaluating the remaining integrals exactly. However, we can investigate the behaviour in the near-horizon limit, equivalent to taking $z_1/a_1,z_2/a_2\rightarrow\infty$. To avoid subtleties associated with the branch cut at infinity, we first performing a $z\rightarrow 1/z$ transform, 
\be
\ba
F(a,b,c;z)&=\frac{\Gamma(c)\Gamma(b-a)}{\Gamma(b)\Gamma(c-a)}(-z)^{-a}F(a,a-c+1,a-b+1,1/z)\\
&+\frac{\Gamma(c)\Gamma(a-b)}{\Gamma(a)\Gamma(c-b)}(-z)^{-b}F(b,b-c+1,-a+b+1,1/z)~,
\ea
\ee
which allows us to expand in the limit where the fourth argument of the hypergeometric function vanishes. Applying this to the product of hypergeometric functions in \eqref{eq:2pttemp}, and then expanding around $z/a\rightarrow \infty$ yields, to first order,
\be
\ba
{}_2F_1&\lp1+i\omega_+,1+i\omega_-,2,\frac{-z_1^2}{a_1^2}\rp{}_2F_1\lp1-i\omega_+,1-i\omega_-,2,\frac{-z_2^2}{a_2^2}\rp\\
=\lp\frac{a_1a_2}{z_1z_2}\rp^2&\left[
\lp\frac{z_2}{a_2}\rp^{2i\omega_-}\frac{\Gamma\lp i(\omega_--\omega_+)\rp}{\Gamma(1+i\omega_-)\Gamma(1-i\omega_+)}
+\lp\frac{z_2}{a_2}\rp^{2i\omega_+}\frac{\Gamma\lp-i(\omega_--\omega_+)\rp}{\Gamma(1-i\omega_-)\Gamma(1+i\omega_+)}\right]\\
\times&\left[\lp\frac{z_1}{a_1}\rp^{-2i\omega_+}\frac{\Gamma\lp i(\omega_--\omega_+)\rp}{\Gamma(1+i\omega_-)\Gamma(1-i\omega_+)}
+\lp\frac{z_1}{a_1}\rp^{-2i\omega_-}\frac{\Gamma\lp-i(\omega_--\omega_+)\rp}{\Gamma(1-i\omega_-)\Gamma(1+i\omega_+)}\right]~.\label{eq:Fprod}
\ea
\ee
Without loss of generality, we shall assume $z_2>z_1$. Substutiting this expansion into the two point function yields
\be
\langle0| \Phi(0,a_1,z_1)\Phi(0,a_2,z_2)|0\rangle
=\int\dd\omega_+ \int\dd\omega_- (U+L)
\ee
where we've defined
\be
\ba
U\equiv&\omega_- \omega_+ e^{\pi (\omega_+-\omega_-)} \text{csch}(\pi \omega_-) \text{csch}(\pi \omega_+) \left(\frac{a_1}{a_2}\right)^{-i (\omega_-+\omega_+)}\\
& \left(\frac{\left(\frac{z_1^2}{a_1^2}\right)^{-i \omega_-} \Gamma (i \omega_+-i \omega_-)}{\Gamma (1-i \omega_-) \Gamma (i \omega_++1)}+\frac{\left(\frac{z_1^2}{a_1^2}\right)^{-i \omega_+} \Gamma (i \omega_--i \omega_+)}{\Gamma (i \omega_-+1) \Gamma (1-i \omega_+)}\right) \left(\frac{\left(\frac{z_2^2}{a_2^2}\right)^{i \omega_-} \Gamma (i \omega_--i \omega_+)}{\Gamma (i \omega_-+1) \Gamma (1-i \omega_+)}\right)
\ea
\ee
\be
\ba
L\equiv&\omega_- \omega_+ e^{\pi (\omega_+-\omega_-)} \text{csch}(\pi \omega_-) \text{csch}(\pi \omega_+) \left(\frac{a_1}{a_2}\right)^{-i (\omega_-+\omega_+)}\\
& \left(\frac{\left(\frac{z_1^2}{a_1^2}\right)^{-i \omega_-} \Gamma (i \omega_+-i \omega_-)}{\Gamma (1-i \omega_-) \Gamma (i \omega_++1)}+\frac{\left(\frac{z_1^2}{a_1^2}\right)^{-i \omega_+} \Gamma (i \omega_--i \omega_+)}{\Gamma (i \omega_-+1) \Gamma (1-i \omega_+)}\right) \left(\frac{\left(\frac{z_2^2}{a_2^2}\right)^{i \omega_+} \Gamma (i \omega_+-i \omega_-)}{\Gamma (1-i \omega_-) \Gamma (i \omega_++1)}\right)~.
\ea
\ee

We will first perform the integral over $\omega_-$, by viewing $U$ and $L$ as functions on the complex $\omega_-$-plane. One can then easily show the following:
\begin{itemize}
 \item $U$ and $L$ have simple poles at $\omega_-=\omega_+ \pm n i$, for $n \in \mathbb{Z}$.
 \item $|U(i \omega_-)|\rightarrow 0$ and $|L(-i\omega_-)| \rightarrow 0$ in the limit $\omega_-\gg 1$.
 \item $U+L$ has no poles on the real $\omega_-$ axis.
\end{itemize}
With these properties in hand, the integral can be performed via the residue theorem, where we close the contour in the upper/lower half-plane for $U$/$L$, respectively:
\be
\ba
\langle \Phi \Phi\rangle&=\int\dd\omega_+ \int\dd\omega_- (U+L)\\
&=\int\dd\omega_+ 2\pi i\left[- \text{Res}(L,\omega_+)+\sum_{n=1}^\infty \left( \text{Res}(U,\omega_++ni) - \text{Res}(L,\omega_+-ni)\right) \right]\\
&\approx \int\dd\omega_+ \frac{-2}{\pi} \left(\frac{z_1 }{z_2} \right)^{-2i\omega_+}\left(2\log\left(\frac{a_2}{z_2}\right)+2 \gamma + \psi(i\omega_+)+\psi(-i\omega_+) \right)
\ea
\ee
where $\psi(z)=\Gamma'(z)/\Gamma(z)$ and $\gamma=-\psi(1)$. In evaluating the residues we used that $z_i/a_i\gg 1$ for $i=1,2$.

The integral over $\omega_+$ is evaluated in a similar fashion. Viewing the integrand as a function in the complex $\omega_+$-plane, one can see that it is well behaved on the real axis, and goes to zero at $+i\infty$. Closing the integration contour in the upper half-plane, the residue theorem yields
\be
\ba
\langle \Phi \Phi\rangle
&\approx \int\dd\omega_+ \frac{-2}{\pi} \left(\frac{z_1 }{z_2} \right)^{-2i\omega_+}\left(2\log\left(\frac{a_2}{z_2}\right)+2 \gamma + \psi(i\omega_+)+\psi(-i\omega_+) \right)\\
&=2 \pi i \sum_{n=1}^\infty \left(\frac{-2 i z_1^{2n}}{\pi z_2^{2n}} \right)
\propto \frac{z_1^2}{z_2^2-z_1^2}
\ea
\ee
which one can recognise as the correct two-point function for a massless scalar in $AdS_{2+1}$, i.e. 
\eq{
\langle \Phi(0,a_1,z_1)\Phi(0,a_2,z_2)\rangle=\frac{1}{e^S \sinh(S)}\propto\frac{z_1^2}{z_2^2-z_1^2}
}
where the geodesic distance $S$ in the near-horizon limit is given by $S=\log\lp z_2/z_1\rp$.

\end{appendix}

\bibliographystyle{arXiv}
\bibliography{Precursors}
\end{document}